%% file: ics13.tex
\def\sharedaffiliation{%
\end{tabular}
\begin{tabular}{c}}

\documentclass[9pt]{sig-alternate}

\usepackage{graphicx,listings,fixltx2e,array,color,algpseudocode}
\usepackage[lofdepth,lotdepth]{subfig}
\usepackage{stmaryrd}
\usepackage[usenames,dvipsnames]{xcolor}

\definecolor{lightgray}{rgb}{0.96,0.96,0.96}

\begin{document}



\title{Locality Optimization for Data Parallel Programs}
\subtitle{NYU CS Technical Report TR2013-955}
\numberofauthors{3}
\author{
\alignauthor Eric Hielscher\\
\email{hielscher@cs.nyu.edu}
\alignauthor Alex Rubinsteyn\\
\email{alexr@cs.nyu.edu}
\alignauthor Dennis Shasha\\
\email{shasha@cs.nyu.edu}
\sharedaffiliation
\affaddr{Computer Science Department}
\affaddr{New York University}
\affaddr{New York, NY, 10003}
}
%
\newcommand{\Map}{\textbf{Map}}
\newcommand{\Reduce}{\textbf{Reduce}}
\newcommand{\Scan}{\textbf{Scan}}
\newcommand{\AllPairs}{\textbf{AllPairs}}
\newcommand{\TiledMap}{\textbf{TiledMap}}
\newcommand{\TiledReduce}{\textbf{TiledReduce}}
\newcommand{\TiledScan}{\textbf{TiledScan}}
\newcommand{\concat}{\ensuremath{+\!\!\!\!+\,}}

\maketitle

\begin{abstract}
Productivity languages such as NumPy and Matlab make it much easier to implement data-intensive numerical algorithms. However, these languages can be intolerably slow for programs that don't map well to their built-in primitives.

In this paper, we discuss locality optimizations for our system Parakeet, a just-in-time compiler and runtime system for an array-oriented subset of Python. Parakeet dynamically compiles whole user functions to high performance multi-threaded native code.  Parakeet makes extensive use of the classic data parallel operators \Map, \Reduce, and \Scan.  We introduce a new set of data parallel operators, \TiledMap, \TiledReduce, and \TiledScan, that break up their computations into local pieces of bounded size so as better to make use of small fast memories.  We introduce a novel tiling transformation to generate tiled operators automatically.  Applying this transformation once tiles the program for cache, and applying it again enables tiling for registers.  The sizes for cache tiles are left unspecified until runtime, when an autotuning search is performed.  Finally, we evaluate our optimizations on benchmarks and show significant speedups on programs that exhibit data locality.

\end{abstract}




\section{Introduction}
\label{sec:intro}
Productivity languages such as NumPy~\cite{Oliphant07} and Matlab~\cite{Matlab10} make it much easier to implement data-intensive numerical algorithms. However, these languages rely on  library functions to attain acceptable performance and can become intolerably slow for programs which don't map well on to precompiled primitives. In this paper, we discuss our system Parakeet~\cite{Rubi12}, a just-in-time compiler and runtime system for an array-oriented subset of Python. Parakeet dynamically compiles whole user functions to high performance multi-threaded native code. This paper focuses on our use of data parallel operators as a basis for locality enhancing optimizations. We present a high-level syntactic transformation which enables tiling for better use of cache and registers. 

The data parallel model allows programmers to expresses algorithms by creating and transforming collections using high level constructs. For example, whereas an imperative language would require an explicit loop to sum the elements of an array, a data parallel language might instead implement summation via some form of reduction operator. The enduring appeal of data parallel constructs lies in the flexibility of their semantics. A data parallel transformation only specifies what the output \textit{should be}, not how it is computed. This makes data parallel programs amenable to parallelization (as the name suggests), both in terms of coarse-grained data partitioning and fine-grained SIMD vectorization. 

In our system, we transform programs written with the classic data parallel operators ($\Map$, $\Reduce$, and $\Scan$) to process small local pieces of data at a time. To express this locality we have introduced three new data parallel operators: $\TiledMap$, $\TiledReduce$, and \TiledScan. These operators are not exposed to the programmer but rather are automatically generated by a tiling transformation. For programs which can benefit from spatial or temporal locality of memory access, applying this transformation can result in signficant performance gains. 

In summary, the contributions of this paper are:
\begin{itemize}
 \item A novel code transformation that tiles data parallel programs to improve both cache and register usage. 
 \item An online autotuning search to select tile sizes.
 \item A just-in-time optimizing compiler for an array-oriented DSL embedded in Python, which can readily outperform naive C implementations on a variety of tasks. 
\end{itemize}

\section{The Parakeet DSL and Compiler}
\label{sec:parakeet}

Using Parakeet can be as simple as calling a Parakeet library function from within existing Python code. For example, the first call to \lstinline!parakeet.mean(matrix)! will compile a small program to efficiently average the rows of a matrix in parallel. Repeated calls will not incur further compilation costs. If a Python function is wrapped with the \lstinline!@parakeet.jit! decorator, then its body will be parsed by Parakeet and prepared for later compilation. When such a function is finally called, its untyped syntax will be specialized for the types of the given arguments and then compiled and executed. For example, consider the simple function given below:

\begin{lstlisting}[label=lst:add1, caption={Simple Parakeet function}, belowskip=0.5em]
@parakeet.jit
def add1(x):
  return x+1
\end{lstlisting}

If \lstinline!add1! is called with an integer argument, then it will be compiled to return an integer result. If, however, \lstinline!add1! is later called with a floating point input then a new native implementation will be compiled that computes a floating point result. 

This example does not make use of any data parallel operators. In fact, it is possible to generate code with Parakeet using only its capacity to efficiently compile loops and scalar operations. However, even greater performance gains can be achieved through either the explicit use of data parallel operators or, commonly, the use of constructs which implicitly  generate data parallel constructs. For example, if you were to call \lstinline!add1! with a vector, then Parakeet would automatically generate a specialized version of the function whose body contains a $\Map$ over the elements of $x$. This can also be written explicitly: 

\begin{lstlisting}[label=lst:add1, caption={Add 1 to every element}, belowskip=0.5em]
@parakeet.jit
def add1_map(x):
  return parakeet.map(lambda xi: xi +1, x)
\end{lstlisting}

In addition to its core data parallel operators $\Map$, $\Reduce$, and $\Scan$, Parakeet also supports derived data parallel operators. For example, the generalized outer product $\AllPairs$ is ultimately translated into to a nested pair of maps, but can be more covenient to use. 

Functions which contain data parallel operators are aggressively optimized and executed across multiple cores. Aside from the tiling transformation described later, we also perform operator fusion~\cite{Wadler81}, as well as more standard compiler optimizations such as common subexpression elimination, loop unrolling, and scalar replacement. 
Parakeet generates native code using the LLVM compiler infrastructure~\cite{Latt02}.  The Parakeet source is available for download at {\tt http://github.com/iskandr/parakeet}.  We are working toward an official, publicized release in the next few months.

\begin{figure}[!ht]
\centering
\begin{tabular}{| m{1.3cm}m{0.2cm}m{5.6cm} |}
\hline
 & & \\
 expression & ::= & Const \\ 
            & $|$ & $x$\\
            & $|$ & $e_1 \odot e_2 $\\
            & $|$ & $[e_1, \ldots, e_n]$ \\
            & $|$ & $e_1[e_2]$\\
            & $|$ & $\lambda x_1,\ldots,x_n.$block \\ 
            & $|$ & $\Map_\alpha(f, e_1, \ldots, e_n)$\\
            & $|$ & $\Reduce_\alpha(f, \oplus, e_\textrm{init}, e_1, \ldots, e_n)$\\
            & $|$ & $\Scan_\alpha(f, \oplus, f_\textrm{emit}, e_{\textrm{init}}, e_1, \ldots, e_n)$\\
 statement  & ::= & $x$ \textbf{=} $e$\\
            & $|$ & \textbf{return} $e$\\
            & $|$ & \textbf{if} $e_\textrm{cond}$ \textbf{then} block$_\textrm{true}$ \textbf{else} block$_\textrm{false}$\\
            & $|$ & \textbf{for} $x$ \textbf{in} $e_\textrm{seq}$ block$_\textrm{body}$\\
 block      & ::= & statement$^+$\\
 & &\\
\hline
\end{tabular}
\caption{Parakeet's Internal Representation}
\label{fig:parakeet_syntax}
\end{figure}

\section{Tiled Operators}
\label{sec:tiled_operators}
When the data access pattern of a program involves significant locality -- temporal, spatial, or both -- this enables a number of different performance optimizations.  Temporal locality enables much better data cache behavior, as accesses to a data item after the first can result in relatively cheap cache hits as opposed to expensive reads from RAM.  Spatial locality can also improve cache performance, as data is stored in caches in units called cache lines that are typically on the order of 16 words of memory.  When data is accessed in a pattern that uses entire cache lines at a time, all but the first read to an item in the line is serviced by a cheap cache hit.  Locality is also important for good use of processor registers.  It is often very beneficial for performance when data is reused repeatedly in the inner loop of a computation to load a small amount of data into registers and then to perform the inner loop on the registers.  This way, every access after the first involves using a register as opposed to a trip to slower levels of the memory hierarchy.

In order to provide a convenient single abstraction for dividing data parallel operator computations into locality-friendly pieces, we introduce \emph{tiled data parallel operators}, one for each regular data parallel operator.  Tiled data parallel operators are a natural generalization of data parallel operators that, rather than applying their nested function to each element of their input arrays, instead break their input arrays up into groups of elements of bounded size called tiles and execute their nested functions on these tiles.  Users never program directly with tiled data parallel operators -- they are strictly internal syntax for use in locality optimizations.  The Parakeet compiler automatically generates them from untiled code via a tiling transformation described later in Section \ref{sec:tiling_transformation}.

\begin{figure}[!t]
\begin{tabular}{ll}
$X$: & \includegraphics[scale=0.3]{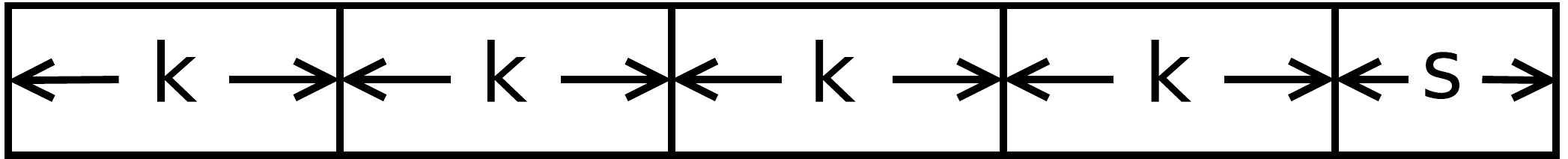}\\[0.4em]
Tiled $X$: & \includegraphics[scale=0.3]{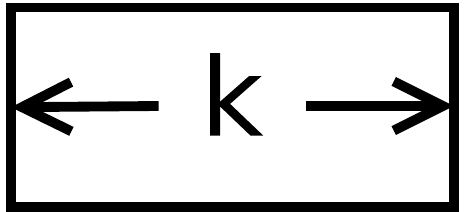}$_0$,\includegraphics[scale=0.3]{TilePiece.pdf}$_1,\dots,$\includegraphics[scale=0.3]{TilePiece.pdf}$_n$,\includegraphics[scale=0.3]{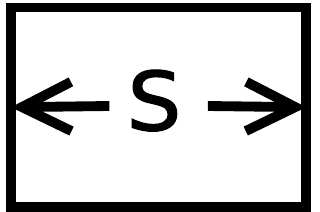}
\end{tabular}
\caption{Decomposition of an Array Into $n$ $k$-length Tiles Plus An $s$-length Straggler Tile}
\label{fig:tiledarray}
\end{figure}

\begin{figure}[!t]
\begin{itemize}
 \item \TiledMap($f$, $f^k$, $X$):\\
   $f^k($\includegraphics[scale=0.3]{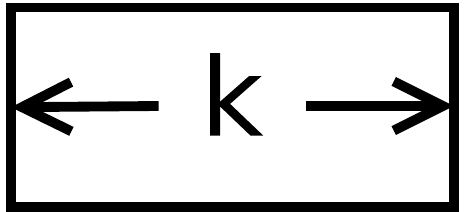}$_0) \concat \dots \concat f^k($\includegraphics[scale=0.3]{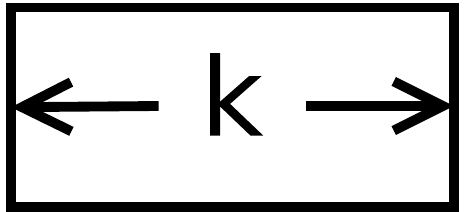}$_n)$ \concat $f($\includegraphics[scale=0.3]{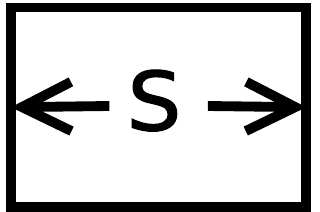})
 \item \TiledReduce($f$, $f^k$, $\oplus$, $X$):\\
 $f^k($\includegraphics[scale=0.3]{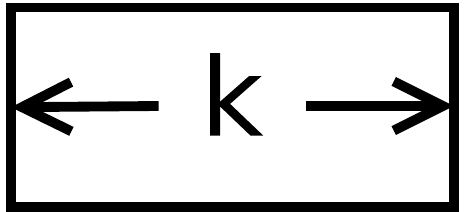}$_0)\oplus\dots\oplus f^k($\includegraphics[scale=0.3]{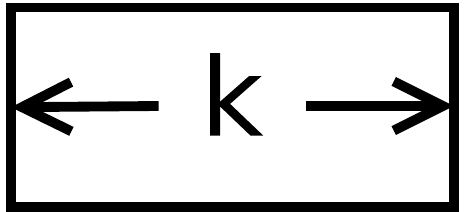}$_n)\oplus f($\includegraphics[scale=0.3]{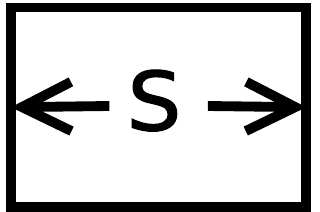})
 \item \TiledScan($f$, $f^k$, $\oplus$, $X$):\\
 $result_0 := f^k($\includegraphics[scale=0.3]{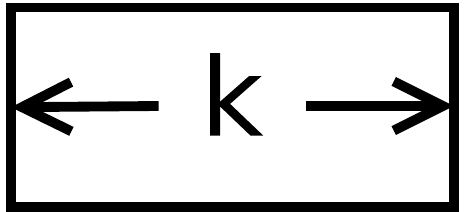}$_0)$\\
 $result_{i=1..n} := $ last el. of $result_{i-1} \oplus f^k($\includegraphics[scale=0.3]{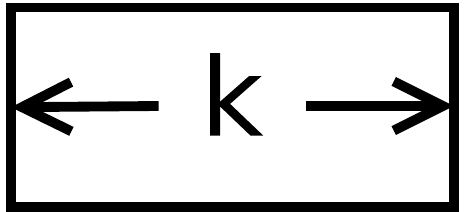}$_i)$\\
 $result_{n+1} := $ last el. of $result_n \oplus f($\includegraphics[scale=0.3]{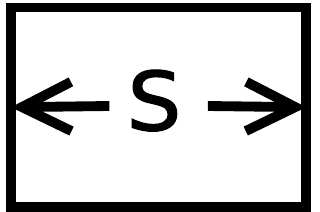})\\
 \textbf{return} $result_0 \concat \ldots \concat result_{n+1}$
\end{itemize}

\caption{Visual Semantics of Tiled Data Parallel Operators}
\label{fig:tiledoperators}
\end{figure}

Figure \ref{fig:tiledarray} shows the decomposition of an array $X$ into $n$ tiles each of length $k$, with an additional tile of length $s$ that contains the leftover elements of $X$ if its length isn't evenly divisible by $k$.  Figure \ref{fig:tiledoperators} gives a visualization of the semantics of tiled data parallel operators.  Tiled data parallel operators can take multiple arguments and axes just like regular data parallel operators, but we omit them from this picture for clarity.  For example, a \TiledMap~decomposes its input arguments into tiles, executes its nested function on each tile (including the straggler tile), and then concatenates the results.  The semantics of \TiledReduce s and \TiledScan s are similarly direct generalizations of their untiled counterparts.

Tiled data parallel operators also take two versions of their nested functions -- one specialized for the specific tile size $k$ (denoted $f^k$), and one generic to array length (simply denoted $f$) which is used to process the straggler tile.  Knowledge that the $f^k$ nested function will be called only on arrays of a fixed size allows Parakeet to optimize it in various ways, for example by removing boundary checks.  In addition, register tiling
optimizes the $f^k$ functions even further.

Decomposing data parallel operators in this way allows the Parakeet compiler to perform both cache tiling and register tiling via the same abstraction.  The decomposition step alone is enough to enable cache tiling, provided that the values of $k$ are chosen properly such that the working set of each function call fits into cache.  When Parakeet tiles for cache locality, the values of $k$ are left undetermined until runtime when an online search is performed to choose good values for them.

By contrast, in order for Parakeet to use tiled data parallel operators to perform register tiling, we fix the $k$ values to small compile-time constants based on a heuristic that takes into account the number of registers on the target machine.  Parakeet then lowers $f^k$ into a loop and completely unrolls it, which is possible due to its fixed length.  Afterward, scalar replacement is applied to remove as many direct memory accesses to the tile's elements as possible, instead keeping them in registers.

\input{tiling_transformation}

\section{Tile Sizes}
\label{sec:online_tuning}

We run our transformation twice, once to general a level of tiles for the L1 cache, and once to generate tiles for registers.  The register tiles are set statically at compile time using a heuristic that attempts to use as many of the registers available on the target machine without exceeding that number.  There are two methods used to set cache tile sizes in the literature: statically via models, and empirically via autotuning.

\subsection{Statically Estimating Tile Sizes}
A body of work exists on attempting to devise models for general tiling or tiling for specific domains that allows for good static setting of tile size parameters~\cite{Cole95, Shir12, Yoto03, Yoto05}.  However, in our experiments we weren't able to get any of the static models to perform as well as a dynamic tile size search, which backs up the continued widespread use of autotuning.  In our setup, we use the DL and ML algorithms from~\cite{Shir12} as low and high initial guesses for tile sizes.  The DL algorithm is designed to provide a pessimistic estimate of cache behavior and thus minimum values for tile sizes, while the ML algorithm provides optimistic maximum tile size estimates.  These algorithms provide a surface of minimum and maximum tile sizes respectively in the tile search space.  We generate square-shaped tiles on each boundary surface and use these as starting points for a dynamic tile size search, described in the next section.

\subsection{Online Autotuning Tile Sizes}
A common algorithm for searching across tile sizes used in the literature is the Parallel Rank Ordering algorithm (PRO), similar in flavor to the Nelder-Mead method~\cite{Taba05}.  In this method, a simplex of tile sizes is maintained with 2 points in the tile space for each tile parameter.  In each step of the algorithm, a reflection through each point of the simplex is evaluated by executing the program with the tile sizes corresponding to the point.  As many such points are evaluated as possible in parallel.  Then the reflection is either accepted or rejected, and further shrinking or expanding of the simplex is potentially done.

\begin{figure}
\centering
\includegraphics[scale=0.4]{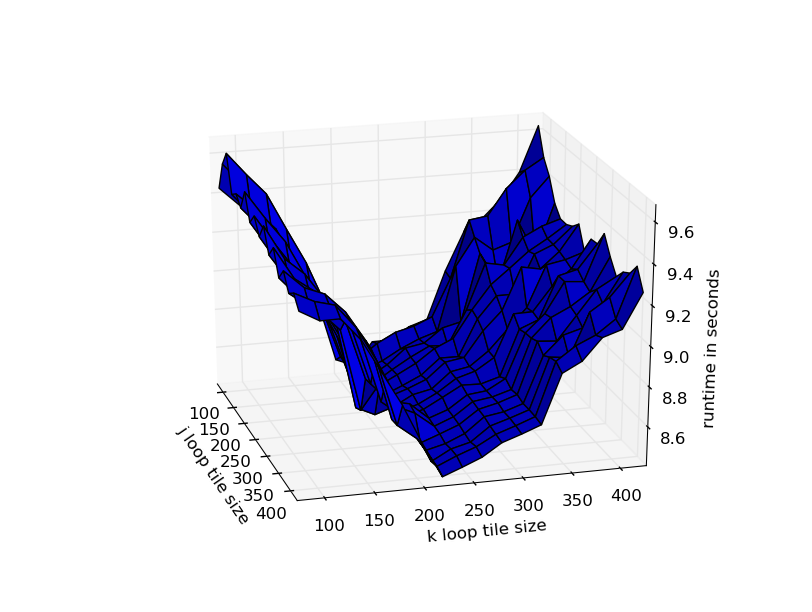}
\caption{Matrix Multiply Performance vs Tile Size}
\label{fig:mm_tile_sizes_bowl}
\end{figure}

In our experiments, we weren't able to get this method to work well enough as it had too high overhead compared to the benefit of reaching better tile sizes.  In our setup, we aren't trying to find the best possible tile size, as we're tuning tiling online for user code rather than tuning it offline for a standard matrix multiplication library.  Thus, we need our algorithm to converge quickly to a fairly good tile size, and then stop searching and exploit the good tile size for the rest of the run.  Each set of tile sizes tested is relatively expensive, as if the sizes are bad they slow down the whole run.  Further, for many of the benchmarks we tested, the performance relative to tile size settings involved a region of tile size settings that had good performance, surrounded by settings where performance was bad (shown for matrix multiplication in Figure \ref{fig:mm_tile_sizes_bowl}).  The performance variation within the good region wasn't high enough to justify further tuning once it was reached.

Thus we opted for a different search algorithm that required fewer samples to reach the good region of the tile size space.  In each time step, we take the current best performing point, initialized to the average of the DL and ML estimates discussed in the previous section.  For each tile size, we take a Gaussian sample with standard deviation equal to the half the difference between these estimates.  A sample for each tile size forms a candidate setting of tiles, and a number of these is evaluated in parallel on different cores of the machine.  If any perform better than the previous best point, they become the new best.  We set heuristics for determining what percentage of the total computation should be spent searching, how long to let each candidate evalutation run, and how many times no change in best point should lead to a termination of the search.

\input{evaluation.tex}

\input{related_work.tex}

\section{Conclusion}
\label{sec:conclusion}
We have presented tiled data parallel operators, novel high-level syntactic constructs for breaking up data parallel programs into locality-friendly pieces.  Our tiling transformation automatically generates tiled versions of programs from programs written using regular operators in Parakeet, a high level array-oriented DSL embedded in Python.  We apply this transformation twice, once to enable cache tiling and a second time to enable register tiling.  Our system includes an autotuner that, after estimating candidate tile sizes using published algorithms, tunes these while the program runs, resulting in a modest performance boost.  We evaluate our optimizations on two benchmarks and show significant performance improvements over untiled code and favorable performance compared to C versions.






\bibliographystyle{abbrv}
\bibliography{Parakeet}{}





\end{document}

%% file: tiling_transformation.tex
\section{Tiling Transformation}
\label{sec:tiling_transformation}
In this section, we present our algorithm for automatically translating a Parakeet function with data parallel operators into a version with tiled data parallel operators.  The basic idea should be intuitive -- we wrap data parallel operators in tiled versions of themselves (say, a \Map~in a \TiledMap) -- and in this way use the tiled data parallel operator to break up the original computation into locality-friendly pieces.  The original data parallel operators become the nested functions of the tiled data parallel operators, as the original computation still needs to be performed on each tile.  However, there are a number of issues that make things somewhat more complicated.

\begin{figure}
\centering
\includegraphics[scale=0.4]{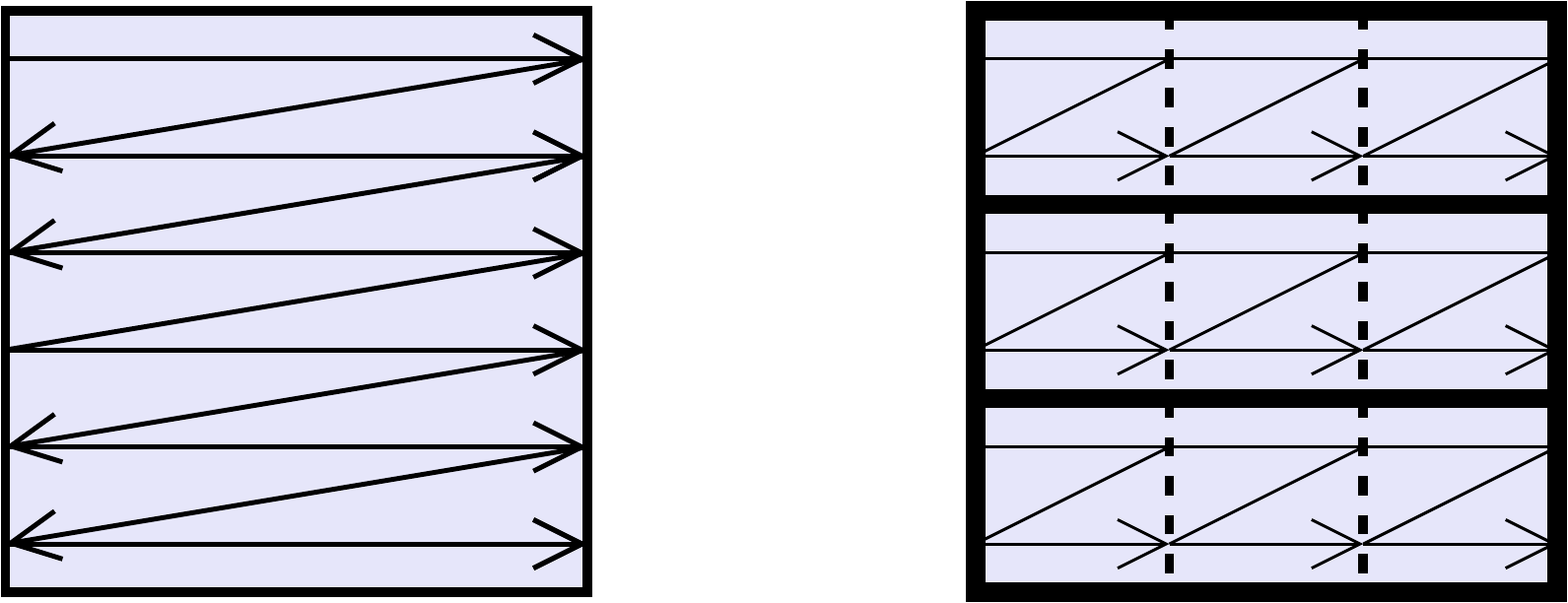}
\caption{Iteration Order of Untiled (left) and Tiled (right) 2D Row Sums}
\label{fig:sum_2d_rows_iterations}
\end{figure}

Let us use summing the rows of a 2D matrix as a simple running example.  Parakeet code for this is given in Listing \ref{lst:sum_2d_rows}.  The iteration pattern through the elements of the array for this version is shown on the left side of Figure \ref{fig:sum_2d_rows_iterations} -- the \Reduce~iterates over each row in its entirety before the \Map~moves on to the next row.

Such a program has the potential to benefit from cache tiling due to the cache line effect discussed in Section \ref{sec:tiled_operators}.  For example, if the data is layed out such that adjacent elements of columns (rather than rows) are adjacent in memory, then whenever an element of the array is read some number of neighboring elements in its column will also be brought into cache as they'll be in the same cache line as the read element.  If the rows are larger than the size of the cache however, these neighboring elements will have been evicted by the time it is their turn to be read.  Thus what could have been a cheap cache hit if the program were tiled instead becomes a costly trip to RAM.  Let us walk through the steps required to tile this program for cache.

\begin{lstlisting}[label=lst:sum_2d_rows, caption={Sum Each Row of a 2D Array}, belowskip=0.5em]
def add2(a, b):
  return a+b

def sum_row(row):
  return Reduce(lambda x:x, row, init=0,
                combine=add2, axes=[0])

def sum_rows(Xs):
  return Map(sum_row, Xs, axes=[0])
\end{lstlisting}

We would like to break up both the \Map~and the \Reduce~by adding tiled versions of them.  To break up the \Map, we wrap the entire computation in a \TiledMap~that groups the rows of the array into tiles.  These row tiles are represented on the right side of Figure \ref{fig:sum_2d_rows_iterations} by the bold boxes around groups of rows.  To tile the \Reduce, we then break these row tiles further via the use of a \TiledReduce~that divides the rows in each row tile into a series of partial rows.  The divisions added by the \TiledReduce~are represented in the figure by the dashed lines.  The iteration order through each of the final tiles is represented on the right side of the figure by the arrows.  Note that this procedure lets us break every dimension of the input into pieces of bounded size.  Thus regardless of the size of any of the dimensions of the input array \lstinline{Xs}, with properly chosen tile sizes we can ensure that the amount of data in the smallest tiles fits into whatever size cache is 
being targeted for optimization.

The final tiled code for summing the rows of a 2D array is given below in Figure \ref{fig:tiledsum2drows}.  Let's break it down piece by piece to explain how our algorithm produces this code.

\begin{figure}[!t]
\begin{tabular}{| l |}
\hline\\
def identity(x): \\
\quad \textbf{return} x \\
\\
def add2(a,b): \\
\quad \textbf{return} a+b\\
\\
def sum\_row(rowTile): \\
\quad\textbf{return} \Reduce(identity, rowTile, init=0,\\
\quad\quad\quad\quad\quad\quad\quad\quad\quad combine=add2, axes=[0])\\
\\
def sum\_rows(XsTile): \\
\quad\textbf{return} \Map(sum\_row, XsTile, axes=[0])\\
\\
def tiledadd2(as,bs): \\
\quad\textbf{return} \Map(add2, as, bs)\\
\\
def tiledsum\_row(XsMapTile): \\
\quad\textbf{return} \TiledReduce(sum\_rows, XsMapTile, init=0,\\
\quad\quad\quad\quad\quad\quad\quad\quad\quad\quad\quad combine=tiledadd2, axes=[1])\\
\\
def tiledsum\_rows(Xs): \\
\quad\textbf{return} \TiledMap(tiledsum\_row, Xs, axes=[0])\\
\\\hline
\end{tabular}
\caption{IR for Tiled Sum Each Row of a 2D Array}
\label{fig:tiledsum2drows}
\end{figure}

\subsection{Visited Operators List}

First, notice that at a high level the code is as we would expect -- a \TiledMap~is contained in the outermost function \lstinline{tiledsum_rows}, whose nested function \lstinline{tiledsum_row} contains a \TiledReduce, whose nested function \lstinline{sum_rows} in turn contains the original computation.  The first thing our algorithm must do is keep track of the nesting of data parallel operators encountered as it steps through a program so as to be able to create a tiled copy of that nesting.

A formalized description of our entire algorithm is given in figures \ref{fig:full_tiling_algorithm_part1}, \ref{fig:full_tiling_algorithm_part2}, and \ref{fig:full_tiling_algorithm_part3}.  In the formal description, the sequence of encountered data parallel operators is represented by $\sigma$.  The generic operation of \emph{tiling} a statement, a block of statements, or a data parallel operator is represented by the $\llbracket\rrbracket$ operator.  Recall that the entry point into Parakeet is always either a data parallel operator or an outermost function called from Python.  Tiling an outermost function involves simply tiling its block of statements.

\begin{figure*}
\centering
\begin{tabular}{| m{0.01cm}m{5.3cm}m{0.5cm}m{7.4cm} |}
\hline
& & &\\
& $\alpha$ & ::= & Axes along which an adverb slices\\
& & & each of its arguments \\
& $\sigma$ & ::= & Ordered sequence of visited adverbs\\
& $\Delta$ & ::= & Maps variables to the list of axes remaining\\
& & & of the original variables of which they\\
& & & are a piece\\
& $\epsilon$ & ::= & Maps variables to the nesting depths\\
& & & at which they were tiled \\
& $\epsilon[x]$ & ::= & $\langle\rangle$ if $x \notin \epsilon$ or if $x$ is a constant\\
& $\text{FV}(e)$ & ::= & The set of free variables in expression $e$\\
& & & \\
& $\text{BuildTree}(\sigma, \epsilon, x_1,\ldots,x_n,$ & &\\
& \ \ \ \ \ \ \ \ \ \ \ \ \ \ \ \ \ \ $block$) & ::= & 
      
   $\sigma_0(\text{vars}_0, \lambda \text{vars}_0.$\\
& & & 
   $\quad\sigma_1(\text{vars}_1, \lambda \text{vars}_1.$\\
& & & 
   $\quad\quad\ldots$\\
& & & $
   \quad\quad\quad \sigma_d(\text{vars}_d,\lambda \text{vars}_d.block$) $\ldots ))$\\
& & & where \\
& & & $ \quad d = |\sigma|$\\
& & & $ \quad \text{vars}_i = \left\{x_j \middle| i \in \epsilon[x_j]\right\}$\\
& & & \\

& $\llbracket f = \lambda x_1,\ldots,x_n.$body$\rrbracket$ & $\longmapsto$ & $f' = \lambda x_1,\ldots,x_n.\text{body}'$\\
& & & where\\
& & & \quad$\sigma = \langle\rangle$\\
& & & \quad$\Delta[x_i] = \langle 0, 1, \ldots, $ rank($x_i)-1\rangle$\\
& & & \quad$\epsilon = \{\}$\\
& & & \quad body$' = \llbracket\text{body}\rrbracket^{\sigma,\Delta,\epsilon}$\\
& & & \\
\hline
\end{tabular}
\caption{Tiling Transformation, Definitions}
\label{fig:full_tiling_algorithm_part1}
\end{figure*}

\begin{figure*}
\centering
\begin{tabular}{| m{0.01cm}m{5.3cm}m{0.5cm}m{7.4cm} |}
\hline
& & &\\
& \multicolumn{3}{l|}{\textit{Statement and Expression Transformations}}  \\
& & & \\
& $\llbracket\textbf{\textrm{return}}$ $e\rrbracket^{\sigma, \Delta, \epsilon}$ 
  & $\longmapsto$ 
  & $\epsilon$, \textbf{return} $\llbracket e \rrbracket^{\sigma, \Delta, \epsilon}$\\
& & & \\

& $\llbracket x \textbf{=} e\rrbracket^{\sigma, \Delta, \epsilon}$ 
  & $\longmapsto$ & $\epsilon'$, x \textbf{=} $e'$\\
& & & where\\
& & & $\quad \epsilon'[x] = \bigcup \left\{\epsilon[y] \; \middle| \; y \in \text{FV}(e)\right\}$\\
& & & \quad if $e$ contains an adverb:\\
& & & \quad\quad $e' = \llbracket e \rrbracket^{\sigma, \Delta, \epsilon}$\\
& & & \quad else:\\
& & & $\quad\quad \sigma' = \langle \Map_0, \Map_0, \ldots \rangle \text{ such that }$\\
& & & $\quad\quad\ \ \ \ \ \ \ \ \ |\sigma'| = |\epsilon'[x]|$ \\
& & & \quad\quad $e' = \text{BuildTree}(\sigma', \epsilon', \text{FV}(e),$\\
& & & \quad\quad\ \ \ \ \ \ \ \ \ \ \ \ \ \ \ \ \ \ \ \ \ \ \ \ \ $\langle \textbf{return}\;  e \rangle)$ \\
& & & \\
& $\llbracket block \rrbracket^{\sigma, \Delta, \epsilon}$ 
  & $\longmapsto$ & $block'$\\
& & & where\\
& & & \quad $block' = \langle\rangle$\\
& & & \quad $\epsilon' = \epsilon$\\
& & & \quad for $s \in block:$\\
& & & \quad\quad $\epsilon', s' = \langle\llbracket s\rrbracket^{\sigma, \Delta, \epsilon'}\rangle$\\
& & & \quad\quad $block' = block' \concat s'$\\
& & & \\
& $\llbracket e\rrbracket^{\sigma, \Delta, \epsilon}$ 
  & $\longmapsto$ 
  & $e$ (if $e$ is not an adverb)\\
& & & \\
\hline
\end{tabular}
\caption{Tiling Transformation, Statements}
\label{fig:full_tiling_algorithm_part2}
\end{figure*}

\begin{figure*}
\centering
\begin{tabular}{| m{0.01cm}m{5.3cm}m{0.5cm}m{7.4cm} |}
\hline
& & & \\
& \textit{Adverb Transformations} & &  \\
& & & \\

& $\llbracket \Map_\alpha(v_1,\ldots,v_n, f)\rrbracket^{\sigma, \Delta, \epsilon}$ 
  & $\longmapsto$   
  & $\TiledMap_{\alpha'}$ ($v_1,\ldots,v_n,f^{\uparrow}$)\\
& & & where \\
& & & $\quad x_1, \ldots, x_n = \text{args}(f)$ \\
& & & $\quad d = |\sigma|$ \\ 
& & & $\quad \epsilon'[x_i] = \epsilon[v_i] \concat \langle d\rangle$\\
& & & $\quad \sigma' = \sigma \concat \langle \Map_\alpha \rangle $ \\
& & & $\quad \alpha'_i = \Delta[v_i][\alpha_i]$\\
& & & $\quad \Delta'[x_i] = \Delta[v_i]$ with element \\
& & &  \quad\ \ \ \ \ \ \ \ \ \ \ \ \ \ \ $\Delta[v_i][\alpha_i]$ removed\\
& & & $\quad \text{If } f$ contains adverbs, \\
& & & \quad\quad body$^{\uparrow} = \llbracket \text{body}(f) \rrbracket^{\sigma', \Delta', \epsilon'} $\\
& & & $\quad \text{Otherwise}$, \\
& & & \quad\quad body$^{\uparrow} = \text{BuildTree}(\sigma', \epsilon', x_1, \ldots, x_n,$\\
& & & $\quad\quad\ \ \ \ \ \ \ \ \ \ \ \ \ \ \ \ \ \ \ \ \ \ \ \ \ \ \ \ \ \ \ \ \text{body}(f))$\\ 
& & & $ \quad f^{\uparrow} = \lambda x_1, \ldots, x_n . \text{body} ^ {\uparrow}  $ \\
& & & \\
& $\llbracket \Reduce_\alpha(v_1,\ldots,v_n, f, \oplus)\rrbracket^{\sigma, \Delta, \epsilon}$ 
  & $\longmapsto$   
  & $\TiledReduce_{\alpha'}(v_1, \ldots, v_n, 
    f^{\uparrow}, 
    \oplus^{\uparrow})$\\
& & & where \\

& & & $\quad x_1, \ldots, x_n = \text{args}(f)$ \\
& & & $\quad d = |\sigma|$ \\
& & & $\quad \epsilon'[x_i] = \epsilon[v_i] \concat \langle d\rangle$\\
& & & $\quad \sigma' = \sigma \concat \langle \Reduce_\alpha \rangle $ \\
& & & $\quad \alpha' = \Delta[v_i][\alpha_i]$\\
& & & \quad body$^{\uparrow} = \text{BuildTree}(\sigma', \epsilon', v_1, \ldots, v_n,$\\
& & & $\quad\ \ \ \ \ \ \ \ \ \ \ \ \ \ \ \ \ \ \ \ \ \ \ \ \ \ \ \ \ \ \ \ \text{body}(f))$\\
& & & $\quad f^{\uparrow} = \lambda  x_1, \ldots, x_n.\text{body}^{\uparrow}$ \\
& & & $\quad \sigma'' = \langle \Map_0, \Map_0, \ldots \rangle$\\
& & & $\quad\ \ \ \ \ \ \ \ \ \ \text{ such that } |\sigma''| = d$\\
& & & $\quad c_1,\ldots,c_k = \text{args}(\oplus)$\\
& & & $\quad \l = $ max$_i(\epsilon[v_i])$\\
& & & $\quad \epsilon''[c_j] = \langle 0,1,\ldots,l-1\rangle$\\
& & & $\quad \oplus^{\uparrow} = \text{BuildTree}(\sigma'', \epsilon'',
\text{args}(\oplus),$\\
& & & $\quad\ \ \ \ \ \ \ \ \ \ \ \ \ \ \ \ \ \ \ \ \ \ \ \ \ \ \ \ \text{body}(\oplus) $)\\
& & & \\
& $\llbracket \Scan_\alpha(v_1,\ldots,v_n, f, \oplus)\rrbracket^{\sigma, \Delta, \epsilon}$ 
  & $\longmapsto$ & Same as \Reduce~but with\\
& & & \Scan~and \TiledScan\\

& & &\\
\hline
\end{tabular}
\caption{Tiling Transformation, Adverbs}
\label{fig:full_tiling_algorithm_part3}
\end{figure*}

The rest of the parameters of the \TiledMap~match those of the original \Map~-- it iterates over the variable \lstinline{Xs} on axis 0.  Thus our algorithm can simply copy these parameters from the original \Map~and use them for the tiled version.

Next, let's examine the \lstinline{tiledsum_row} function and its \TiledReduce.  It iterates over the \lstinline{XsMapTile}, analagous to the original \Reduce~iterating over \lstinline{row}, and its initial value of 0 is the same as that of the \Reduce.  Its nested function is the original \lstinline{sum_rows} function that was tiled, which is what we would expect since it is the final tiled data parallel operator in the nesting.  However, its axes and combine parameters require further explanation.

\subsection{Axes of Tiled Operators}

The \TiledReduce's iteration axis for the \lstinline{XsMapTile} is 1, not 0 like in the original \Reduce.  To understand why, look again at the iteration order shown in Figure \ref{fig:sum_2d_rows_iterations}.  Notice that each \lstinline{XsMapTile} is a two dimensional group of rows, whereas the original reduce iterated over single 1D rows.  A key difference between tiled data parallel operators and regular data parallel operators is that tiled data parallel operators always operate on tiles of the outermost arguments that are the same rank as the entire outermost arguments themselves.  In contrast, regular data parallel operators operate on pieces of the arguments with progressively decreasing rank, as each regular operator slices away one dimension.  Thus the axes of iteration for regular operators have a local view on the input arguments to the outermost function -- they only see the dimensions of the arguments that remain after any previous operators have sliced some dimensions away.  This is why, even 
though the \Reduce~in the original code iterates over what is axis 1 of the outermost 2D input argument \lstinline{Xs}, its axis of iteration over its 1D row is 0.

In order to properly update the axes of variables such that they fit the tiled operators, we need to keep around some state for each variable we encounter.  For this we maintain a mapping from variables to the axes at which prior adverbs in the nesting have sliced them.  In the formal description, this mapping is denoted by $\Delta$.  Using these lists of axes, we can map from the local axis of the original adverb back to the global axis of the entire tile/outer variable as required.

\subsection{Tiled Combine Functions}

The combine function of the \TiledReduce~in Figure \ref{fig:tiledsum2drows} also needs further explanation.  Recall that the purpose of a combine function is to provide a way to combine two partial results of a reduction.  For a \TiledReduce, we need the combine function to combine two partial results of executing the nested function on \emph{tiles}.


Let's walk through expanding the \TiledReduce's combine function.  The number of data parallel operators in the visited data parallel operators list $\sigma$ is one adverb prior to this point: the \Map.  Thus we've expanded the arguments to the \TiledReduce's nested function as well as the nested function's return value by 1 rank.  Hence we need to increase the rank of the arguments and results of the \Reduce's combine function by 1 in order to create a combine function suitable for the \TiledReduce.  We do this by calling a helper function we name $BuildTree$.  This function wraps the \Reduce's combine function \lstinline{add2} in one \Map, as shown in Figure \ref{fig:tiledsum2drows}.

The purpose of this \Map~is to ``peel off'' the rank that was added by the previous adverb having been tiled.  This way, we can get to arguments of the rank that the original combine function expects, and then call the original function on those arguments.  By mapping across the extra dimensions added (one in this case), we ensure that we call the combine function on each element of the tiles, thus maintaining the semantics of the original program.

Since there are no more data parallel operators to tile, we are now ready to insert the original computation as the nested function of the final tiled data parallel operator, and the tiling transformation terminates.  The reader following along with the formal algorithm will notice that a \Reduce~or a \Scan~always terminates the tiling algorithm.  This is because it is not safe in general to split the results of one reduction and have these partial results form the input to another one (for example, the partial results of the minimum of an array aren't meaningful).  In addition, notice that the insertion of the original computation in the formal algorithm actually involves another use of the $BuildTree$ function.  This is in order to support nested non-operator statements, as described next.

\subsection{Nested Statements}

One last issue not covered in the summing rows example is what happens in the presence of statements that don't include data parallel operators inside functions being tiled (for example, scalar operations, indexing into variables, or control flow).  In the case of the outermost function, these statements are simply left alone.  However, inside a nested function of a tiled operator, we must be careful to maintain the semantics of the original program.  If we simply added these statements both to the nested functions of the tiled operators in addition to keeping them in the nested functions of the inner regular operators, this may result in an error.  For example, in the case of indexing into a variable, this would amount to indexing into a tile, and then in the innermost computation, indexing again into a piece of that tile.  This would index once too many times, and break the program's semantics.

To solve this issue, we do two things.  First, we disallow control flow in functions being tiled; if control flow is found, the tiling transformation is undone and the code is left untiled.  Control flow can be handled in these cases by predication~\cite{Bena10}, but we leave this for future work.

Next, we place all of the other non-operator statements only in the nested functions of the tiled adverbs, and remove them from the copy of the original program that includes the original operators.  In the case of indexing, e.g., this is the only legal possibility, as if the indexing were delayed until the inner regular operators executed, portions of the original variables could be present and iterated over that weren't visible the original program.  Note that we only remove statements \emph{that have an adverb in the same scope}.  The innermost computation, which can contain arbitrary non-operator code, is left untouched.

When we get to the last operator in the tree and want to splice in the altered version of the original program, we take the innermost nested function (that contains no operators), and call the $BuildTree$ function on its body.  This function doesn't wrap the body in \Map s as in the case of tiled combine functions.  Instead, it wraps the block in the tree of operators from the original program.  Thus we create a tree with the original operators, but devoid of any non-operator statements.

One thing we must ensure, however, is that the proper variables are passed as arguments to each operator in this tree.  For this we keep track of the \emph{list of nesting depths} at which each variable was an argument to a data parallel operator.  In the formal algorithm, this additional mapping is denoted by $\epsilon$.  Often variables are passed into nested functions as closure arguments for use in deeper levels of a program's nesting.  We don't want to alter the behavior of the program by adding these as arguments to the operators.  Hence for each operator we need to add we look up in the $\epsilon$ mapping which free variables of the innermost block were arguments to an operator at that nesting depth.  This allows us to add the proper arguments to each operator, and we pass the other needed arguments through as closure arguments to the operator's nested function.

Finally, we must tile the non-operator statements as well.  This is because they are present in tiled functions, and so will receive tiles as arguments instead of those of the original ranks.  To do this, we compile a list of the statement's free variables, and from this a list of all nesting depths at which these variables were arguments to adverbs.  We then call the $BuildTree$ function on the statement to wrap it in a number of \Map s equal to the number of such nesting depths.  We use the depths in the same way as in building the inner computation to pass the correct variables to each \Map.  This has the effect of peeling off the extra ranks as discussed in the case of tiled combine functions.

This method of dealing with scalar statements has the potential to be wasteful in that it generates array temporaries when originally there were none.  It has the benefit of making the algorithm simpler as roughly the same unpacking logic can be applied to all cases of extra ranks due to tiling.  If many of these expanded scalar statements exist in a function, data parallel operator fusion is able to combine them all into a single tree of \Map s, mitigating some of this cost.  An alternative would be to keep the statements for which it is safe (such as scalar operators) in the inner functions of the tiled computation so as not to generate the temporaries.  We leave this for future work.

%
%

%% file: evaluation.tex
\section{Evaluation}
\label{sec:evaluation}

We evaluate our cache and register tiling optimizations on K-Means Clustering and Matrix Multiplication. We have implemented a number of different programs in Parakeet -- for example, Gaussian blurring and $O(N^2)$ N-Body simulations -- but must omit their results for lack of space.

We evaluate our benchmarks on a system with an Intel i7 960 3.2GHz processor and 16GB of RAM.  This processor has 4 hyperthreaded cores, each with a 32KB L1 data cache with 64 byte cache lines.  Parakeet reads these hardware characteristics from the {\tt /proc} and {\tt /sys/devices} filesystems and uses them to configure both the cache tiling and register tiling optimizations.  

In all of these performance results, register tiling and cache tiling were turned on and off together.  In the interest of space, and since we don't vary any parameters for register tiling across any of the runs, we don't present separate numbers for register tiling.  It was very substantial for performance however, accounting for 60\% of the speedups due to tiling on average.  Also, aside from those in the section on Compile Time, no performance results include compile times.

\subsection{Matrix Multiplication}
\label{sec:matrix_multiplication}

In this section, we present results for a 2D matrix multiplication benchmark, the actual Parakeet code for which is in Listing \ref{lst:parakeet_mm}.  Note that \AllPairs~is Parakeet syntactic sugar for two nested \Map s each of which iterates over one of the arguments to the \AllPairs.  Note that for these numbers, we pre-transposed the matrices such that they are both laid out for the best performance.  This is actually the worst-case scenario for our tiling optimizations, as they would provide a substantially larger (roughly 12X) performance boost when run on improperly laid out data.
\\
\begin{lstlisting}[label=lst:parakeet_mm, caption={Parakeet Matrix Multiply}, belowskip=0.5em]
def dot(x, y):
  return x*y

def mm(Xs, Ys):
  return AllPairs(dot, Xs, Ys)
\end{lstlisting}

\subsubsection{Overall Performance}

\begin{figure*}
\centering
\subfloat[][Runtimes with varying rows in left-hand matrix]{
\includegraphics[scale=0.4]{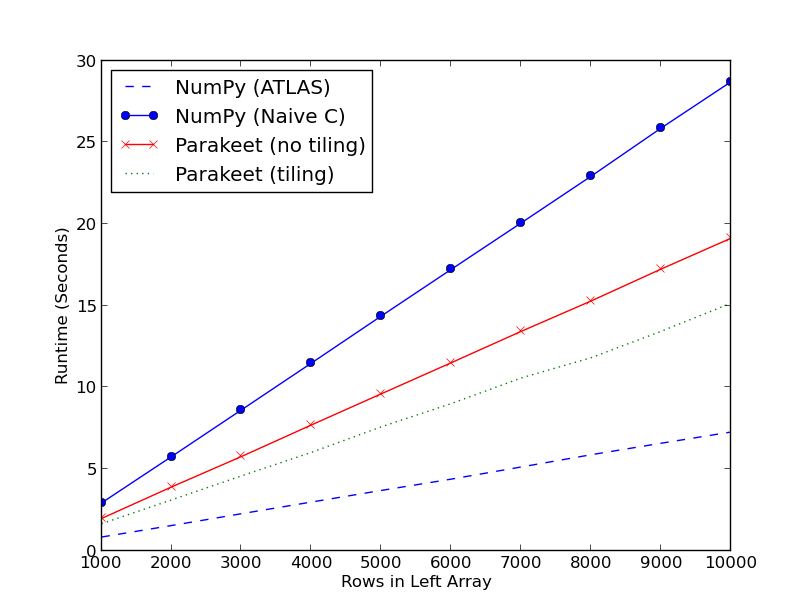}
\label{fig:mm_perf_varying_xrows}}
\quad
\subfloat[][Runtimes with varying lengths of rows]{
\includegraphics[scale=0.4]{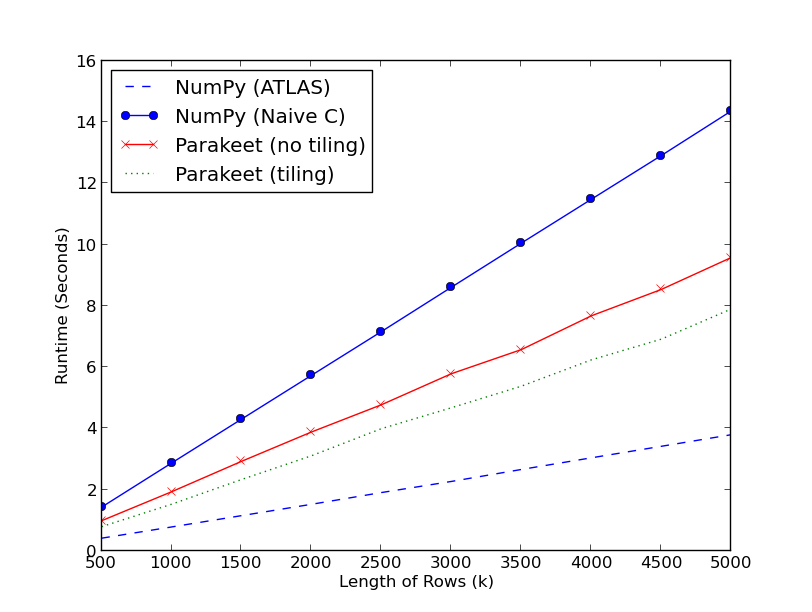}
\label{fig:mm_perf_varying_rowlengths}}
\caption{Matrix Multiply Runtimes}
\label{fig:mm_perf}
\end{figure*}

In Figure \ref{fig:mm_perf}, we present Parakeet's performance on matrix multiplication as compared against NumPy's.  NumPy uses an implementation of BLAS, a standard linear algebra interface, to perform matrix multiplications.  We configure NumPy to use two different BLAS versions -- one written as naive C consisting of three nested \lstinline{for} loops; and a stock ATLAS~\cite{Whal00} implementation that comes with Ubuntu Linux.  For optimal performance, one needs to do a lengthy autotuning of ATLAS for a particular machine.  We use the stock version as we simply want to provide a rough reference point for calibrating the meaning of our results.  In each of these graphs, we used our online autotuner to find good cache tile sizes.

On the left-hand side of the figure,
the number of rows in the left-hand matrix vary along the X axis.
The length of rows and the number of rows in the right-hand matrix are held fixed 3000.  In this figure, we see that our tiling optimization improves performance between 26.3\% and 30.8\% over not tiling. 
On the right-hand side of the figure, we fix the number of rows in both matrices to be 3000 and vary the length of the rows.  Here, the tiling speedup varies between 21.1\% and 24.8\%.  In both cases,
Parakeet is around 2 times slower than our ATLAS version, which recall has been heavily hand-optimized.

\subsubsection{Autotuning Performance}

In Figure \ref{fig:mm_autotuning}, we break down the performance of the autotuner by showing Parakeet times both with the autotuning search as well as Parakeet times using the fixed tile sizes found in a previous search.  We compare these with using the fixed tile sizes for the DL and ML algorithms from~\cite{Shir12}, as well as the performance when using the average of the DL and ML estimates as the fixed tile sizes.

By comparing the difference between the runtime with the search and with cached tile sizes, we see that the overhead of performing the search is very small.  The time to switch between different tile sizes during a search is close to 0, and so any overhead is almost entirely due to the penalty from running worse tile sizes than those that are eventually found.  We also see that, while the autotuner leads to the best runtimes especially on larger data sizes, the performance boost it adds over the ML estimates in particular isn't very large.  On average, the autotuner increases performance around 2.3\% for all benchmarks and data sizes we tried.  While this is only a modest performance boost, it is consistent, and if there are any other programs for which the tile estimates perform badly, the autotuner should be able to increase performance even more by finding better tiles.

\begin{figure*}
\centering
\begin{tabular}{|l|c|c|c|c|c|}
\hline
Array Size & Parakeet (with search) & Parakeet (cached tile sizes) & DL sizes & ML sizes & DL\&ML Avg\\\hline
3000x3000 & 4.58s & 4.51s & 5.03s & 4.55s & 4.61s \\\hline
10000x3000 & 15.16s & 15.11s & 16.89s & 15.33s & 15.51s \\\hline
10000x500 & 2.57s & 2.57s & 2.99s & 2.67s & 2.72s \\\hline
\end{tabular}
\caption{Matrix Multiply Autotuner Performance}
\label{fig:mm_autotuning}
\end{figure*}

\subsubsection{Performance vs. Other Compilers}

\begin{figure}
\includegraphics[scale=0.4]{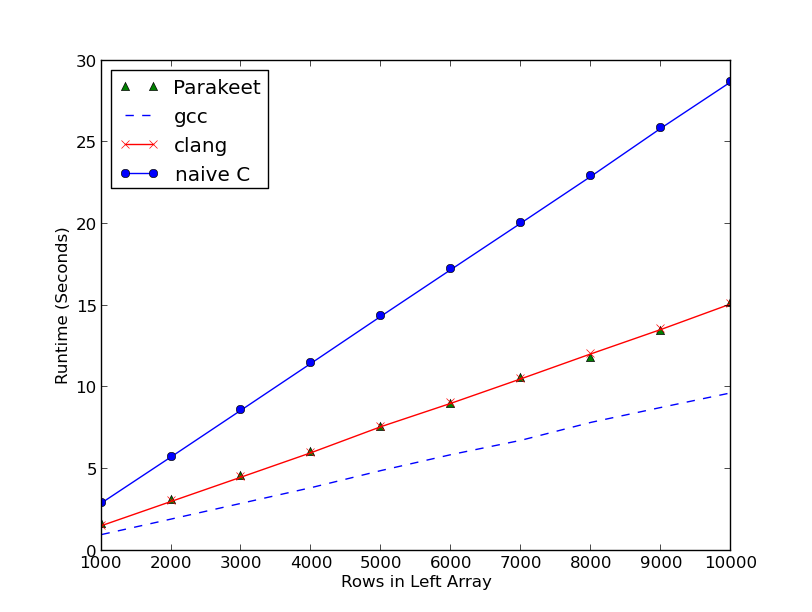}
\caption{Matrix Multiply Performance Compared To C}
\label{fig:mm_vs_c}
\end{figure}

To provide a test of Parakeet's performance relative to other compilers, we compare Parakeet's performance to a hand-optimized C version with manual cache and register blocking compiled with both gcc and clang (the LLVM C compiler), including all relevent optimization flags.  We also include a naive \lstinline{for} loop C version with only the \lstinline{-O3} flag for reference.  All of these versions were launched on 8 threads on the Parakeet runtime's backend.  The results are shown in Figure \ref{fig:mm_vs_c}.  

First, notice that Parakeet's performance roughly matches that of clang, while both are roughly 1.5X slower than gcc.  We take this as evidence that we are approaching a performance wall due to the underlying performance of LLVM.  We discovered that roughly half of gcc's relative performance gain over clang is due to gcc having a better vectorizer.  We hope that in the future, we can either add a vectorizer to Parakeet or that LLVM's vectorizer will improve.


\subsection{K-Means Clustering}
\label{sec:kmeans_clustering}

We present results for K-Means Clustering in Figure \ref{fig:kmeans_perf}.  Here we see that Parakeet is dramatically faster than NumPy, with almost a 4X performance improvement.  The performance benefit of tiling on this benchmark is only 4\% on average however.  This is due to it spending a lower percentage of its computation in operations with much data reuse, and the length of its rows (equal to k, or the number of centroids) being smaller than in much of the matrix multiply benchmark.

\begin{figure}
\centering
\begin{tabular}{|c|c|c|c|}
\hline
Number of & NumPy & Parakeet & Parakeet \\
Data Points &       & (No Tiling) & (Tiling) \\\hline
10000 & 48.01s & 12.91s & 12.46s \\\hline
12500 & 59.95s & 16.05s & 15.51s \\\hline
15000 & 71.83s & 19.16s & 18.51s \\\hline
17500 & 83.78s & 22.28s & 21.40s \\\hline
20000 & 97.61s & 25.53s & 24.45s \\\hline
22500 & 107.62s & 28.54s & 27.35s \\\hline
25000 & 119.56s & 31.66s & 30.41s \\\hline
27500 & 131.57s & 34.82s & 33.37s \\\hline
30000 & 143.32s & 37.89s & 36.42s \\\hline
\end{tabular}
\caption{K-Means Performance with k = 1000, 500 features, and 10 iterations}
\label{fig:kmeans_perf}
\end{figure}

\subsection{Compilation Time}
\label{sec:other_eval}

In Figure \ref{fig:compilation_time}, we give the Parakeet compilation times for the benchmarks without tiling, with only cache tiling, and with both cache and register tiling.  Cache tiling adds a modest amount of compilation time, while register tiling adds around 1 second for each benchmark.  We believe that the extra loop unrolling accounts for most of this added time.  Our compiler was written entirely in Python and we haven't spent any effort optimizing its compile times, so we are hopeful that these numbers can be brought down in the near future.  In addition, we imagine a common use case for Parakeet to be repeated calls to a Parakeet function inside a large numerics computation.  In these cases such compile times would only contribute a tiny fraction of overall program runtimes.

\begin{figure}
\centering
\begin{tabular}{|c|c|c|}
\hline
 & Matrix Multiply & K-Means \\\hline
No Tiling & 0.19s & 0.10s \\\hline
Cache Tiling & 0.27s & 0.31s \\\hline
Cache Tiling + & & \\
Register Tiling & 1.26s & 1.29s \\\hline
\end{tabular}
\caption{Parakeet Compilation Times}
\label{fig:compilation_time}
\end{figure}

%% file: related_work.tex
\section{Related Work}
\label{sec:related_work}
Our work builds upon a range of existing fields of study including optimization of data parallel and array programs, just-in-time compilation, loop optimizations, analytic performance modeling, and autotuning.  Our system is unique in that it is the first to unify: high level, dynamic array languages, JIT compilation, high level tiling, and online autotuning into one system.

The first language to feature data parallel abstractions was APL~\cite{Iverson62}, whose central programming constructs involved high-level manipulation of n-dimensional arrays. The eminent parallelizability of the language's core operators inspired early research in vector processors ~\cite{Thurber70} and parallelization~\cite{Lincoln70}. As computers with massively parallel hardware became more common in the 1980s, many languages such as C~\cite{Hatc91}, Fortran~\cite{Fox90}, and Lisp~\cite{Steele86} were retrofitted with data parallel extensions. More recently, data parallel constructs have appeared repeatedly as core primitives for high level languages and libraries which compile to FPGA descriptions~\cite{Gokhale93}, GPU programs~\cite{Tard06,Cata11}, and even the coordination of distributed computations~\cite{Yu08}.


There have been numerous projects that accelerate high level languages, be it via high performance library calls as in NumPy~\cite{Dubo96}; via JIT compilation as in LuaJIT~\cite{PallLu}, MaJIC~\cite{Alma02}, and the fledgling Numba project~\cite{ContNu}; or via dynamic compilation to C++ as in Copperhead~\cite{Cata11}.  Parakeet lies in some sense in between Numba and Copperhead, but .

There has been extensive research on loop optimizations, including cache tiling, register tiling, data copying, data padding, and loop unrolling optimizations (e.g.~\cite{Lam91, Wolf91a}).  Much of this work uses the polyhedral model, a sophisticated technique that transforms the iteration space of a loop nest via algebraic manipulation~\cite{Bond08, Wolf91a}.  This includes work on combining analytic models for selecting tile sizes with offline autotuning~\cite{Pouc10} and work on on generating parameterized tiles for imperfectly nested loops~\cite{Hart09}.

There have been a host of analytic models for determining parameter settings for dense matrix multiplication~\cite{Cole95, Yoto03, Yoto05}.  Wolf et al.~developed an analytic model for use in a compiler to determine tiling and loop unrolling settings statically for sequential C and Fortran programs~\cite{Wolf96}.  Recent work on analytical bounds for optimal tile sizes~\cite{Shir12}.

In recent years, offline autotuning has emerged as the accepted best practice for optimizing numerical code~\cite{Asan06}.  Libraries such as ATLAS for dense linear algebra~\cite{Whal00} and FFTW for Fourier transforms~\cite{Frig05} deliver the best performance available across a wide range of architectures and platforms for their specific problem domains via an extensive offline search performed at installation time.  Other autotuning systems include Chill~\cite{Chen05b} and Active Harmony~\cite{Tiwa11}.